# Heliostat Optical Error Inspection with Polarimetric Imaging Drone

Mo Tian[1], Kolappan Chidambaranathan[1], Md Zubair Ebne Rafique[1], Neel Desai[1], Jing Bai[1], Randy Brost[2], Daniel Small[2], David Novick[2], Julius Yellowhair[3], Yu Yao[1,*]

[1] Electrical, Computer and Energy Engineering, Arizona State University. Tempe, AZ

2 Sandia National Laboratories. Albuquerque, NM

3 Gryphon Technologies. Albuquerque, NM

* Corresponding Author. Email address: yuyao@asu.edu

## Abstract

On a Concentrated Solar Power (CSP) field, optical errors have significant impacts on the collection efficiency of heliostats. Fast, cost-effective, labor-efficient, and non-intrusive autonomous field inspection remains a challenge. Approaches using imaging drone, i.e., Unmanned Aerial Vehicle (UAV) system integrated with high resolution visible imaging sensors, have been developed to address these challenges; however, these approaches are often limited by insufficient imaging contrast. Here we report a polarimetry-based method with a polarization imaging system integrated on UAV to enhance imaging contrast for in-situ detection of heliostat mirrors without interrupting field operation. We developed an optical model for skylight polarization pattern to simulate the polarization images of heliostat mirrors and obtained optimized waypoints for polarimetric imaging drone flight path to capture images with enhanced contrast. The polarimetric imaging-based method improved the success rate of edge detections in scenarios which were challenging for mirror edge detection with conventional imaging sensors. We have performed field tests to achieve significantly enhanced heliostat edge detection success rate and investigate the feasibility of integrating polarimetric imaging method with existing imaging-based heliostat inspection methods, i.e., Polarimetric Imaging Heliostat Inspection Method (PIHIM). Our preliminary field test results suggest that the PIHIM hold the promise to enable sufficient imaging contrast for real-time autonomous imaging and detection of heliostat field, thus suitable for non-interruptive fast CSP field inspection during its operation.

## 1. Introduction

Concentrated Solar Power (CSP) plants are designed to use central towers to collect light that is reflected and refocused from heliostat mirrors for power generation and storage. Commonly, a CSP field consists of one or more central towers surrounded by hundreds to over 100,000 heliostats (e.g., the Ivanpah Solar Power Facility). Depending on their location and distance from the tower, these heliostats are designed with appropriate tracking orientation and mirror curvatures to provide the best focusing efficiency. Heliostats in Concentrated Solar Power (CSP) systems are operated with electro-mechanical drive systems and supported by adjustable structures such as metal bars and rings. Over time, these components may experience gradual misalignment due to mechanical wear, material fatigue, or environmental factors. Additionally, sudden deviations in accuracy can occur as a result of extreme weather events, such as strong winds, temperature fluctuations, seismic activity, or other external disturbances. The difference between the designed ideal surface normal vectors of the heliostat mirrors and the actual surface normal vectors is defined as optical error. Optical error is a critical factor that influences both efficiency and safety. [2, 3]. Over the years, different methods have been developed to inspect the heliostat field [4, 5], including photogrammetry [6, 7], Hartman type methods [8], fringe- and deflectometry-based techniques (structured light reflection [9–11], deflectometry [12]), flux mapping [13], optical modeling [14-16], backward-gazing method [17], and machine learning based methods [18]. While multiple heliostat inspection techniques exist, each has specific limitations and areas for improvement. These include the complexity of system implementation [6-8, 12-14], the demand for high computational power [15, 16, 18], limitations in precision [6, 7, 17], and sensitivity to environmental conditions [9-12]. Given these challenges, UAV-based inspections have emerged as a promising alternative due to their flexibility, accessibility, and cost-effectiveness. Compared to traditional inspection techniques, UAV-based methods demonstrate great potential for efficiency and scalability, making them a valuable direction for further research and optimization. In recent years, the Unmanned Aerial Vehicle (UAV) scanning approach for heliostat inspection has shown potential in achieving fast and non-intrusive inspection of the field. These methods mainly rely on the reflection image of certain objects into the target mirrors, where the difference between captured images and images calculated with optical model can be used to infer the optical errors of the heliostat mirror facets. For example, the nonintrusive optical (NIO) method utilizes the reflection image of the tower in the CSP field [19, 20]. More recently, UAV-based approach, HelioPoint, has also been demonstrated as a fast airborne calibration method for heliostat fields [21]. The Universal Field Assessment, Correction, and Enhancement Tool (UFACET) was proposed and is still under development now [16]. UFACET is an optical error inspection method based on HFACET [14] (Heliostat Focusing and Canting Enhancement Technique). Instead of setting the camera at a fixed position like HFACET, UFACET uses an imaging drone to scan the field and determine the optical errors of multiple heliostats quickly with machine vision. However, both NIO and UFACET methods use a conventional visible camera to detect the edge of the heliostats and mirror facets, which could be challenging in a number of scenarios where the reflection of the mirrors shows similar intensity and color with its surroundings and thus the imaging contrast in visible images is too low for reliable and real-time edge detection. Our previous work demonstrated that polarimetric imaging can enhanced imaging contrast for heliostat mirror edge detection [1].

In this study, we significantly advanced the polarimetric imaging-based inspection method by developing a comprehensive simulation model to estimate changes in the polarization state, while also enhancing data acquisition and processing capabilities. The model provides critical insights into the evolution of polarization states under varying operational conditions, optimizing UAV flight trajectories for more precise data collection. By leveraging polarization imaging, our approach enhances contrast at mirror edges and minimizes environmental interference, improving both the reliability and effectiveness of polarization-based imaging techniques. Furthermore, this method enables more efficient and adaptive flight planning for large-scale CSP field inspections. The Polarimetric Imaging Heliostat Inspection Method (PIHIM) was applied to field test at Sandia National Solar Thermal Test Facility (NSTTF) and the results of optical errors were evaluated and compared to ideal field on-axis canting angles for the heliostats to gain insights of accuracy and limitations [1]. Further improvements on the polarimetric imaging system such as better sensor resolution, a more robust UAV system and better ground truth baselines are required to enhance the accuracy of the PIHIM method.

## 2. Polarimetric Imaging UAV Setup

There are three fundamental properties of light as an electromagnetic wave: intensity, wavelength and polarization. While human eyes can attain the information of intensity and wavelength as brightness and color, we cannot directly see polarization, which describes the geometrical orientation of electromagnetic wave oscillation. Conventionally, we use Stokes Parameters to describe the polarization state of light. Stokes Parameters are composed of four parameters that are usually combined into a 4x1 vector called Stokes vector, describing the total intensity (first term), linear polarization components (second and third term) and circular polarization components. The last term of the Stokes vector is related to circular polarization, which is not discussed or detected in this work. The polarimetric imaging sensor is capable of capturing the intensity information of the four gratings at 0°, 45°, 90° and 135° [22]. The polarimetric imaging sensor (Lucid Triton TRI050S-PC) used in our setup has a resolution of 5.1 million pixels (2048 x 2448), with each pixel measuring 3.45 µm × 3.45 µm. It has a dynamic range of 71 dB, a signal-to-noise ratio of 39 dB, and quantum efficiency >30% in the visible spectrum (400–700 nm). The absolute sensitivity threshold is 10.9 nLux. Attached is a 35mm lens from Kowa. Fig. 1d shows the pixel array of the polarization imaging sensor. Each super-pixel is composed of four pixels with different wire-gratings, and thus the first three terms of the Stokes parameters can be calculated from the four intensity readings. Since we are only considering the linearly polarized components, we can calculate the first three Stokes Parameters [23] as Equation (1):

$$\begin{bmatrix} S_0 \\ S_1 \\ S_2 \end{bmatrix} = \begin{bmatrix} \frac{(I_0 + I_{90}) + (I_{45} + I_{135})}{2} \\ I_0 - I_{90} \\ I_{45} - I_{135} \end{bmatrix} \tag{1}$$

Angle of Polarization (AoP) and Degree of Linear Polarization (DoLP) can be calculated using Equation (2) and Equation (3). The Degree of Linear Polarization (DoLP) describes the fraction of light intensity that is linearly polarized, indicating the strength of polarization. The Angle of Polarization (AoP) specifies the orientation angle of the linear polarization with respect to the sensor’s reference axis.

$$AoP = \frac{1}{2} tan^{-1}\left(\frac{S_2}{S_1}\right) \qquad (2)$$

$$DoLP = \frac{\sqrt{S_2^2 + S_1^2}}{S_0} \qquad (3)$$

Notice that the definition of axis matters in the definition of AoP. In this work, the AoP is defined over a range of 0° to 180° relative to the 0° transmission axis of the polarization filter array in the image plane. The 0° reference corresponds to the horizontal transmission axis of the sensor and is not related to the spatial coordinates or the camera normal.

To realize the in-situ inspection, we built a polarimetric imaging drone for fast scanning of the field [1]. We enhanced both the hardware and software components to enable stable wireless communication, accurate flight logging, and optimized image capture for optical error evaluation. Our polarimetric imaging setup is composed of UAV, a microcontroller and a polarization camera. Fig. 1a shows the polarimetric imaging drone in flight operation at the Sandia NSTTF and Fig. 1b shows a 3D model of polarimetric imaging UAV setup's camera and micro-controller's unit. Our customized graphical user interface (GUI) shown in Fig. 1c allows the operator to view the live feed of the camera at 15 frames per second and capture raw images with the remote control and can also provide an instant preview of the processed image. An adaptive exposure algorithm is running in real time to ensure the captured images are not overexposed in different angles and lighting conditions. Video signals are relayed via HDMI between the camera and Jetson, then wirelessly transmitted between the UAV and ground control station for real-time monitoring and remote operation, as shown in Fig. 1e. The camera is equipped with a 35mm lens to enable capturing closer images of heliostats while the drone flies at a height greater than 5 m above the highest point of the heliostats, in accordance with the safety instructions of the CSP field. This system enabled several field tests at Sandia NSTTF, capturing polarization images across various scenarios for contrast enhancement and optical error calculation.

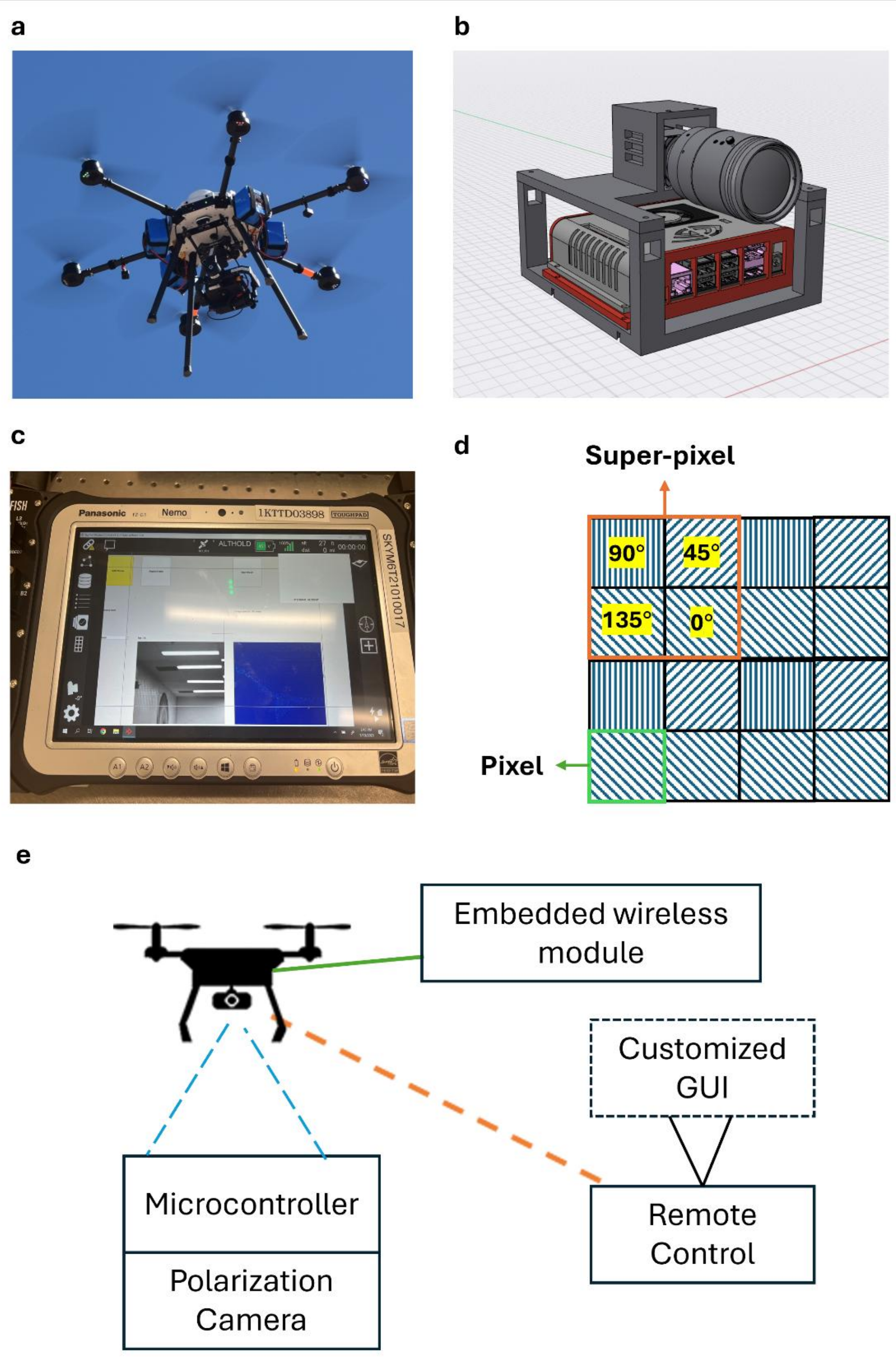

**Figure 1 Polarimetric UAV-system setup and challenging scenarios for conventional visible imaging method. a**, photo of polarimetric imaging UAV setup operating in field test. **b**, 3D model of polarimetric imaging UAV setup's camera and micro-controller's unit. **c**, customized GUI running on the Jetson and viewable from remote control. **d**, polarization sensor's pixel and Super-pixel. **e,** Schematics of system components. **e,** diagram illustrating the UAV setup.

## 3. Polarimetric Imaging Heliostat Inspection Method

Some challenging scenarios regarding edge detection for the conventional imaging UFACET method were found during data collection. We identified the two major ones that can be enhanced by using polarization imaging known as sky-vs-sky and ground-vs-ground scenarios. We found that DoLP images can enhance contrast in the sky-vs-sky scenario, where overlapping edges of two different heliostats, both showing a similar blue color and intensity, are difficult to detect with a visible camera. For the ground-vs-ground scenario when the camera is looking down at a heliostat and capturing the adjacent edge of the ground and the reflection of the ground, AoP images showed good contrast. Here, we have further developed complete models for both scenarios and designed flight tests based on these model simulations. The original UFACET method provides flight paths with constraints to ensure the safety of the flight, speed of the field scanning and capturing the images suitable for analysis. To maximize the benefits of polarimetric imaging, it is essential to simulate DoLP and AoP of the light collected by the camera and use the results as guidance for the flight path design. In total, the ideal flight path to test the polarimetric method will consider three aspects [1]:

1. The waypoints needed to form scenarios for optical error evaluation;
2. The waypoints that collect the reflected light from a skylight region with high DoLP and DoLP gradient, for sky-vs-sky scenario contrast enhancement;
3. The waypoints that collect the reflected light from a skylight region and ground region with significantly different AoP values, for ground-vs-ground scenario contrast enhancement.

An example of the UFACET method is shown in Fig. 2a and Fig. 2b. The back of the Target heliostat forms a reflection image in the Heliostat Under Assessment (HUA) and the difference between the theoretically calculated image and the captured image is used to evaluate the canting error of each facet. Since we are trying to use polarimetric imaging, it is essential to know the incident light's polarization states at different orientations. Based on Rayleigh scattering, the skylight's polarization pattern can be calculated for a specific time, location and angle. Figure 2c and 2d show the DoLP and AoP pattern [24,25] when the sun is at 30° zenith angle (θ) and 235° azimuthal angle ($\phi$). Notice that the values of DoLP and AoP vary at different positions in the spherical coordinates. At Sandia NSTTF, the field coordinates are defined in a Cartesian system where positive x direction is East, positive y direction is North and positive z is up. The origin of the coordinates is at the bottom of the tower (see Supplementary Figure S1). This global coordinate of the field also applies to each individual facet as the origin, with a shift on x-y plane applied. Since the simulation pattern of skylight polarization is only related to the relative position of the Sun, the pattern does not change significantly when the origin is shifted several hundred meters. Thus, for each calculation, the same coordinate system is applied.

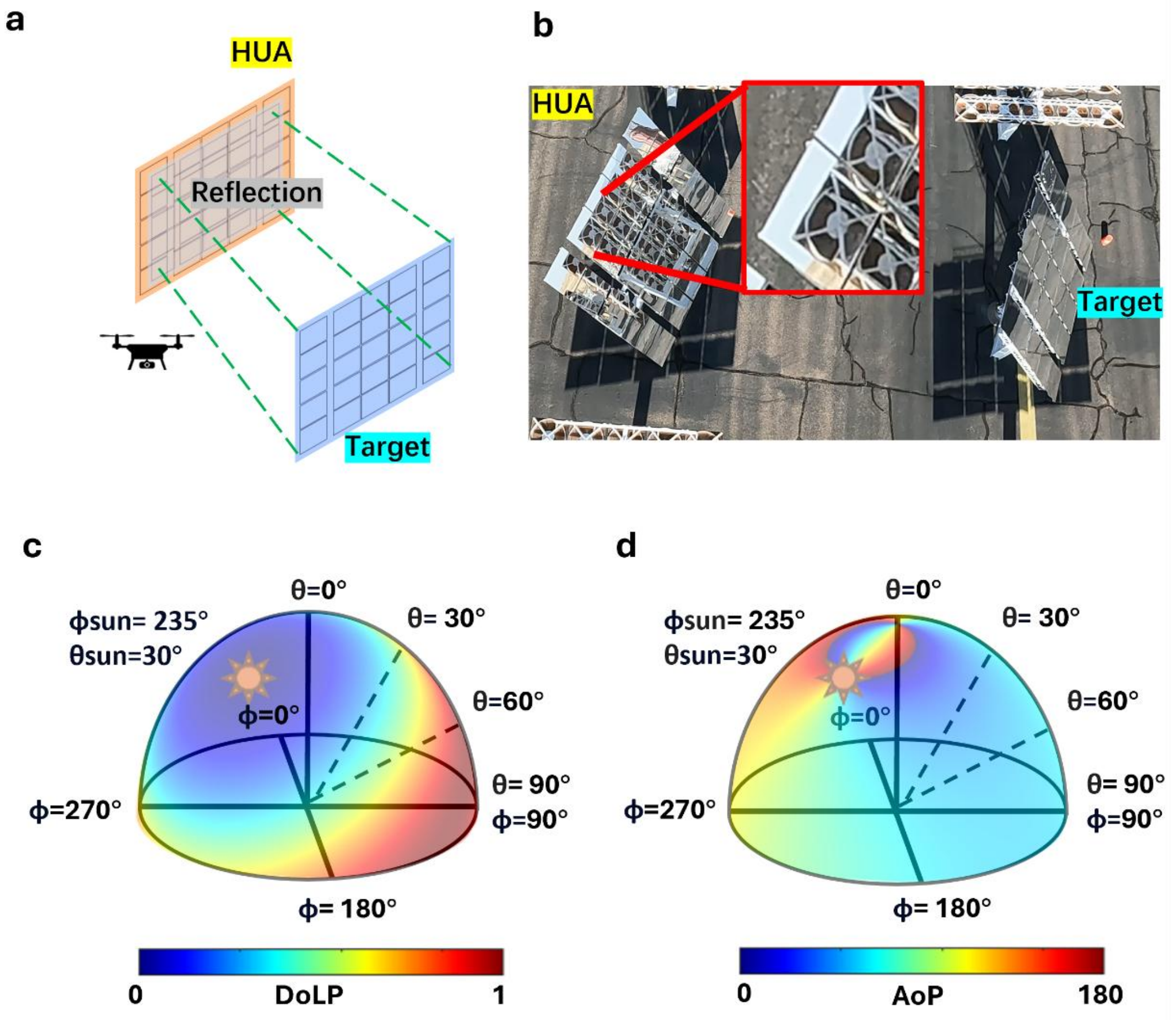

**Figure 2 PIHIM Method overview**. **a**, UFACET image formation using two heliostats. **b,** example image of UFACET and the enlarged features used for optical error evaluation. **c-d**, skylight polarization pattern of DoLP and AoP using Rayleigh scattering model in spherical coordinates.

Furthermore, we need to consider the change of polarization during the light's reflection from a mirror surface or from the ground. The change of polarization during specular reflection off a mirror surface was calculated using Reflection Mueller Matrix derived from Fresnel Laws [23], as in equations (4-7). Here, $S_R$ and $S_I$ denote the Stoke Parameters of reflected light and incident light, respectively. $M_{ab}$ is the reflection Mueller Matrix at interface of media *a* and *b*. $R_l$ and $R_r$ are the parallel and perpendicular reflection coefficients $n_a$ and $n_b$ are the refractive indices of the two materials at the interface, while $\theta_a$ and $\theta_b$ are the incident and refractive angles.

$$S_R = M_{ab} \cdot S_I \qquad (4)$$

$$M_{ab} = \begin{pmatrix} \frac{1}{2}(R_l R_l^* + R_r R_r^*) & \frac{1}{2}(R_l R_l^* - R_r R_r^*) & 0 & 0 \\ \frac{1}{2}(R_l R_l^* - R_r R_r^*) & \frac{1}{2}(R_l R_l^* + R_r R_r^*) & 0 & 0 \\ 0 & 0 & Re\{R_l R_r^*\} & Im\{R_l R_r^*\} \\ 0 & 0 & -Im\{R_l R_r^*\} & Re\{R_l R_r^*\} \end{pmatrix} \qquad (5)$$

$$R_l = \frac{n_b \cos\theta_a - n_a \cos\theta_b}{n_b \cos\theta_a + n_a \cos\theta_b} \quad (6)$$

$$R_r = \frac{n_a \cos\theta_a - n_b \cos\theta_b}{n_a \cos\theta_a + n_b \cos\theta_b} \quad (7)$$

In our previous work, we noticed that with the AoP images it is possible to get very different AoP values on the heliostat reflection and the ground [1]. Here we studied the mechanism thoroughly in this work. The heliostat mirrors are modeled mainly as a smooth and clean surface where the specular reflection can apply. Comparably, the ground has complicated surface morphology and the reflection from it cannot be seen as specular. To model such a rough surface, the Torrance-Sparrow Model [26, 27] can be used as approximation. The rough ground surface is treated as a combination of numerous micro-facets while each of them reflects the incident light individually. As shown in Fig. 3a, the surface normal vectors of these micro facets are treated to follow a certain distribution. For the zenith angle of these surface normal vectors, a half-Gaussian distribution is used for approximation as described in Torrance-Sparrow Model [27]. It is most likely for zenith angle to be 0°, which means that the micro facet is completely flat, and it is least likely that the zenith angle is 90°, meaning that the micro facet is completely vertical to the ground, as indicated in the distribution diagram in Fig. 3b. For the azimuthal angle, since there's no certain preference in direction for the rough ground to face, the horizontal component of the surface normal vector follows a uniform distribution from 0 to 360 degrees. The total reflection result is a normalized integration of all the reflection rays from each micro-facet. The change of polarization during each reflection was calculated using Equations (4-7). In the simulation, the refractive index of the ground was set as 1.45 [28]. The surface was simulated as an ensemble of 100,000 microfacets, and angular sampling was set to 0.5°. These parameters were chosen to balance accuracy with computational efficiency, though both the microfacet density and sampling rate can be further increased for higher-accuracy studies.

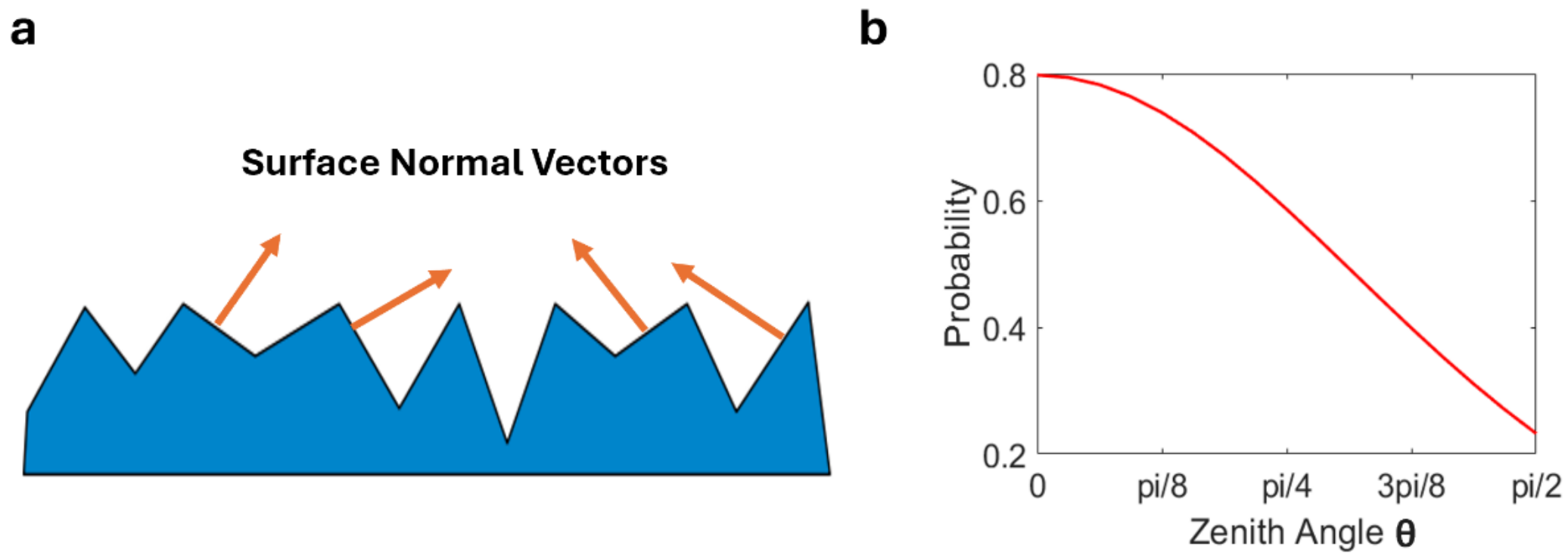

**Figure 3 Modeling rough ground based on Torrance-Sparrow model. a**, rough surface is seen as many micro-facets with different orientations. **b,** Half-Guassian distribution for zenith angle's probability.

When the field is in operation, multiple heliostats focus sunlight on the receiver region, and thus adjacent heliostats point in a similar but not exactly the same direction due to their location differences in the field. We followed the UFACET method to evaluate the canting error of each facet on a heliostat using the back-reflection features of another heliostat [16]. As shown in Fig. 2a UFACET method uses the reflection image of the back of another heliostat to establish the model. In Fig. 2b, the enlarged area shows the details of each single facet's reflection and is used to find the center and edges of the facet image. With known parameters such as camera position, camera viewing angle, calibrated lens and sensor specifications, heliostat locations, heliostat orientations and facet dimensions, it is possible to construct a simulation of the image of these features in the field of view. By comparing the calculated ideal image and the captured image, the optical errors are then evaluated using the difference. To form these images, we design the flight path in the field coordinates such that the camera line of sight is aligned with the reflection of the object heliostat's back. This is an established method that has been tested in the same field. Safety factors such as minimum flight elevation, focusing point of the heliostat reflection and distance to the heliostat mirrors are considered in this part.

## 4. Results and Discussions

### *4.1. Sky-vs-sky Scenario Enhanced by DoLP*

The sky-vs-sky scenario, as shown in Fig. 4a, can be analyzed based on DoLP and DoLP gradient simulation [1]. When we take images with a drone camera, we often encounter a scenario where two adjacent heliostats overlap in the image, but only one of them is the target heliostat of interest. In the UAV scanning method described in UFACET [16], the drone camera position is designed to look at the heliostats with their reflection towards the sky but without direct sunlight. In some situations, two overlapping heliostats both reflect blue sky, resulting in very similar intensity and color, as shown in Fig. 4a. This can result in a low edge contrast between them, causing problems for the detection algorithm to effectively distinguish the border of each heliostat and their facets. Since the skylight forms a natural polarization pattern due to Rayleigh scattering [24], we can use the high DoLP and high DoLP gradient region to enhance the contrast of this scenario [1]. Since these adjacent heliostats are at different locations, their orientations are different as they are tracking the Sun and the collector on the central tower in a CSP field. With a small difference in their orientation angle, they can be reflecting the sky region with rapidly changing DoLP values, and this results in good contrast in DoLP image.

Here, we present a more comprehensive analysis of the field test results in Figure 4, demonstrating the flexibility of this method. Fig. 4a and Fig. 4b show a visible image and a DoLP image, respectively, captured while the drone was flying over the field. The target heliostat pair was set at their tracking position, and all the other heliostats are facing up to simplify the image scene for this test. However, in application this is not necessary, and heliostats should be at their tracking positions while the drone is doing the scanning. Fig. 4c and Fig. 4d are the simulation of the incident light polarization pattern and the heliostat-reflected light polarization pattern, respectively. The five triangles indicate the angle of incident light for five captured images on the same pair of heliostats. The five data points in Fig. 4e and Fig. 4f are corresponding to the triangles indicated in Fig. 4c and Fig. 4d. “Camera Position” 1 to 5 correspond to orange, purple, cyan, blue, and green triangle, respectively. The DoLP values of the front and back heliostats can be derived from the captured images. Since the data was acquired from multiple pixels, there are small local variations of these DoLP values.

In Fig. 4e, the DoLP values and the corresponding standard deviation are shown for comparison. The difference of DoLP between front and back heliostat is the source of the contrast in the polarization images. Further, from the simulation shown in Fig 4d, the DoLP difference between front heliostat and back heliostat can be calculated. The simulation results of difference between front heliostat and back heliostat are then compared with the measured DoLP difference, as shown in Fig. 4f. In general, the data from measurement shows lower DoLP contrast compared to the simulation, which is possibly because of the overall decrease of incident light DoLP due to increased scattering events under various weather conditions [25, 29]. From different camera positions on the same pair of heliostats, we can conclude that the region where DoLP imaging can be applied to the sky-vs-sky scenario is flexible and does not impose significant constraints on the flight. Although it is ideal to utilize the skylight region with the highest DoLP and DoLP gradient, these optimal angles may conflict with other field operation constraints, such as UAV elevation safety requirements or areas obstructed by the central tower. In such cases, the incident light may not exhibit the highest DoLP, yet the method remains effective because the critical factor is the difference in DoLP values between heliostats, rather than the absolute value. While high DoLP incidence is preferred for achieving the best contrast and signal-to-noise ratio, it is acceptable to have decreased DoLP values as long as they are still significantly distinguishable from the background (difference large than 0.2). Considering the field tests carried out at Sandia NSTTF, this condition is mostly achievable between 9:30 AM to 11:30 AM and 2:00 PM to 4:00 PM during the daytime as the location of the Sun during these times is favored for waypoints design. The favored time range can be different for different field location on Earth, time of the year and tracking position of the heliostats.

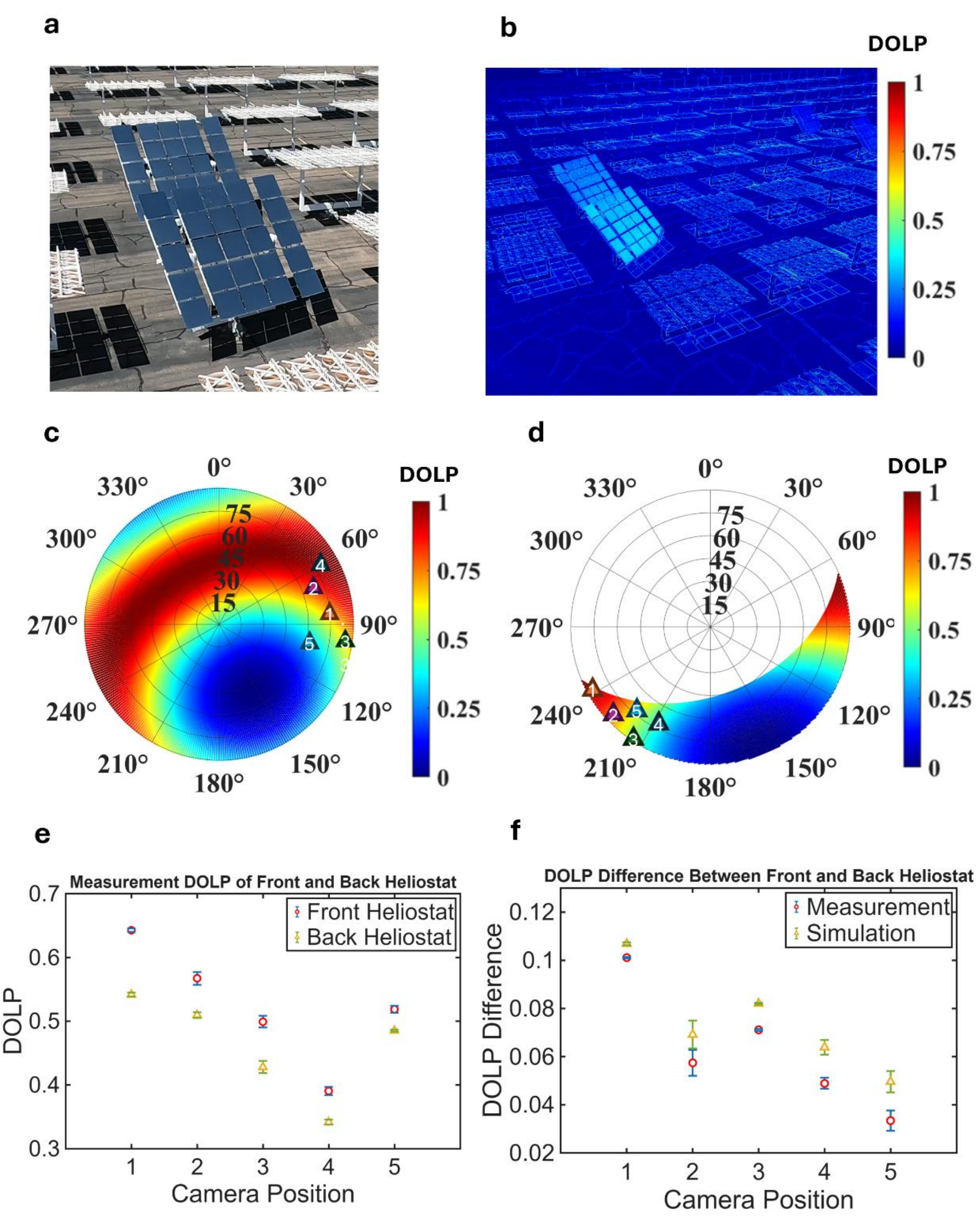

**Figure 4 Results analysis for sky-vs-sky scenario**. **a**, example visible image of the UAV-based polarimetric imaging flight test for sky-vs-sky scenario. **b,** example DoLP image of the UAV-based polarimetric imaging flight test for sky-vs-sky scenario. **c,** simulation of the incident light's DoLP pattern. The five markers indicate the corresponding positions of the skylight input for five different images captured by the polarimetric imaging system looking at the same heliostat pair. **d,** simulation of the reflected light's DoLP pattern. **e,** DoLP values comparison between front and back heliostats overlapping at the edge. These results are from the five images captured as positioned in (d). Camera position indicating different images used for the data shown in this diagram. **f,** the simulated DoLP difference between front and back heliostats and the measured difference acquired from the captured images.

### 4.2. Ground-vs-ground Scenario Enhanced by AoP

Overlapping heliostats is not the only difficulty encountered during overhead inspection with a visible camera drone. When the drone is at a high elevation angle and looks down on the heliostats, the ground can be seen reflected in the heliostats image. When the ground reflection is next to the real ground, the contrast of the bottom mirror edge can be significantly limited due to the similar color. We observed previously that the contrast between the heliostat reflection and real ground can be enhanced using AoP images [1]. Now, based on the Torrance-Sparrow Model we developed, we pre-calculated the desired positions and orientations for the camera to capture images with good AoP difference. Fig. 5a and Fig. 5b show the visible and AoP image of an example ground-vs-ground scenario. Although the bottom edge is challenging for conventional imaging, the AoP shows significant contrast. Based on the calculations of diffuse reflection of the ground using Torrance-Sparrow model and specular reflection of the mirror surface, the simulation plots in Fig. 5c (ground reflection) and Fig. 5d (mirror reflection) provide guidance to find the correct camera position for AoP difference. Fig. 5e is the summary for seven different images taken for the same heliostat in ground-vs-ground scenario. Here the measurement or simulation difference indicates the absolute value of difference in AoP between the heliostat reflection and its adjacent ground reflection. The exact values of these angles and differences are listed in Table 1. For the measured values, the AoP difference was obtained by selecting a 200 × 100 pixels region on the mirror-reflected ground and a corresponding 200 × 100 pixels region on the adjacent actual ground, then subtracting the average AoP of the actual ground from the average AoP of the reflected ground. This averaging approach minimizes local variations and provides a consistent, representative contrast value.

The simulation results are different from the measured results due to the approximation of the model and inaccuracy of the angle. However, in general, while the AoP difference between ground and heliostat is above 50° in absolute value, the contrast is significant statistically in data and visually in the processed figure. Thus, we conclude that the approximation in this Torrance-Sparrow model is sufficient for this purpose. This model was also validated with handheld polarimetric imaging setup and ground (see details in Supplementary Figure S2).

The following assumptions limit the accuracy of this model:

1. The ground is composed of only a single material with the same refractive index.
2. The micro facets only cause specular reflection. The scattering caused by the rough edges and dust on the surface are ignored.
3. The orientation of these facets strictly follows the probability distribution described. In reality, this distribution is more likely skewed due to the artificial formation of the ground as it is a rough paved surface.
4. The Monte-Carlo simulation integrates the reflected light into each 1-degree angle range, ignoring the decimals. This way it can converge faster with less data points calculated.

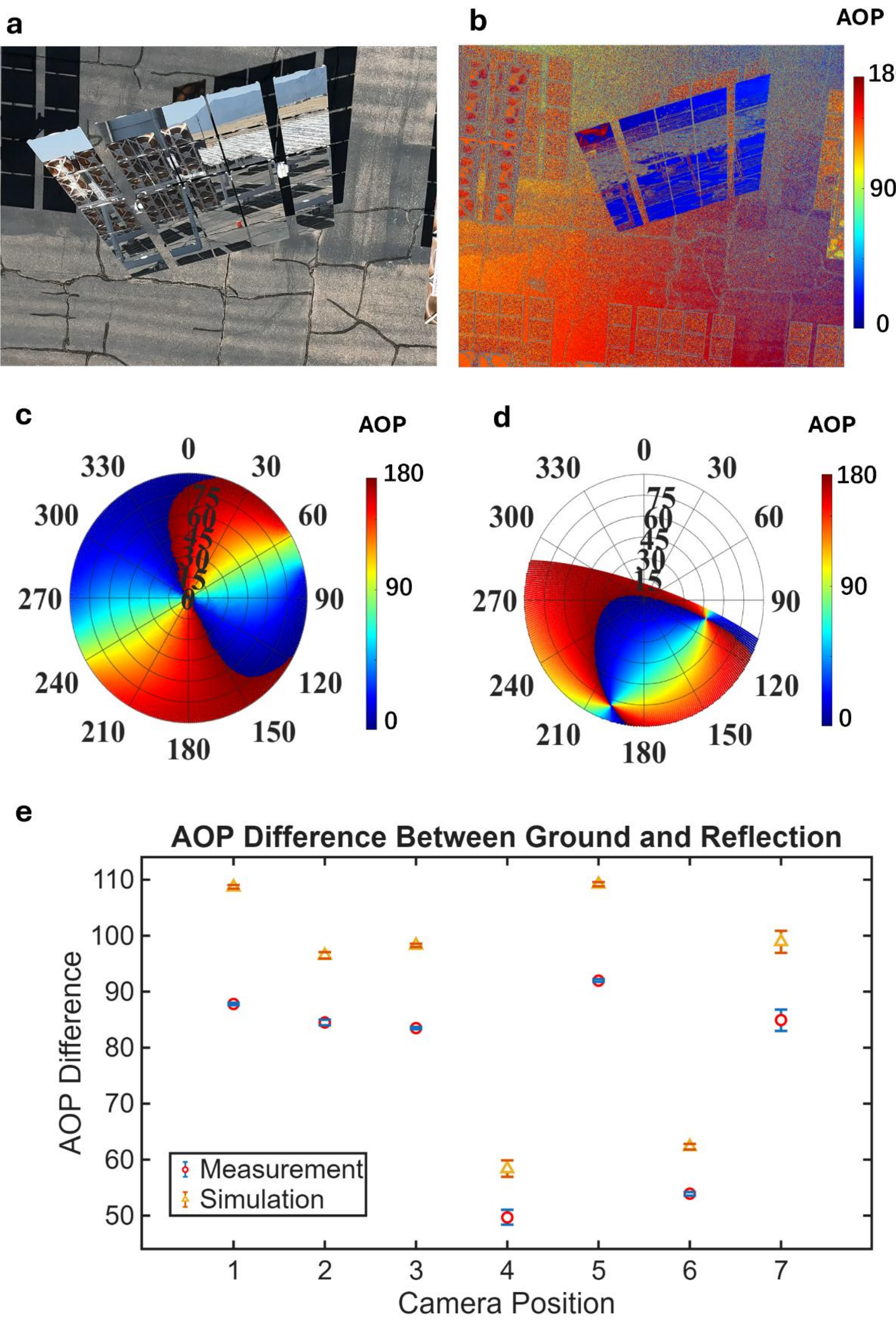

**Figure 5 Results analysis for Ground-vs-ground scenario**. **a**, image captured by visible camera showing ground-vs-ground scenario. **b**, image captured by polarization camera showing ground-vs-ground scenario. **c**, simulations of AOP reflected from the ground. **d**, simulations of AOP reflected from the heliostat. **e**, the simulated AoP difference between ground and heliostat reflection and the measured AoP difference

acquired from the captured images. Camera Position indicates different images used for data.

However, the goal of this simulation is to find regions for the heliostat mirror and ground to have AoP difference on the scale of 50 to 90 degrees. The errors between simulation differences and measured differences are at 10-20 degrees of AoP, making it acceptable for this approach. Here, we define AoP difference as the absolute difference in AoP between the heliostat reflection and its adjacent background (such as the ground). A larger AoP difference corresponds to improved separability of the two regions in polarization space, thereby facilitating edge detection. With this model, it is possible to calculate the flight path for the highest AoP difference while doing a scanning or applying a similar method. Applying the criteria to the previous results [1], the success rate of distinguishing the reflection image and the real ground in ground-vs-ground scenarios is improved from 89% to 96.7%. However, since heliostat inspection methods like UFACET have more constraints to achieve the accuracy required in each field, the model will need more detailed modification in the future for better accuracy.

| Azimuth Angle (°) | Zenith Angle (°) | Measured AoP Difference (absolute value) | Simulation AoP Difference (absolute value) |
|---|---|---|---|
| 272 | 32 | 87.8 | 107.2 |
| 169 | 21 | 84.5 | 94.6 |
| 241 | 33 | 83.5 | 97.3 |
| 114 | 27 | 49.7 | 58.1 |
| 125 | 31 | 92 | 108.3 |
| 189 | 29 | 53.9 | 62.7 |
| 143 | 21 | 84.9 | 97.5 |

**Table 1 Camera angle and AoP difference between ground and heliostat reflection for ground-vs-ground scenario**

### *4.3. Edge Detection*

To validate the benefit of polarization contrast for heliostat inspection, we applied Sobel edge detection to selected regions of interest. Figure 6 shows the Sobel results applied to visible images from Fig. 4a and Fig. 5a, as well as DoLP images from Fig. 4b and AoP images from Fig. 5b.

In the sky-vs-sky case, conventional RGB channels show almost no measurable contrast (Michelson contrast $\approx 1$) at the region of interest, whereas the DoLP images provide significant improvement (average contrast ratio 5.93 from five DoLP images at different angles in Fig. 4). This enables robust edge detection where visible images fail. As shown in Fig. 6a and 6c, Sobel applied to visible images produces fragmented and noisy edges dominated by background interference, while the DoLP image shows clear and continuous edge lines that outline the heliostat. We can also observe from Fig. 6b that the visible image was still able to detect the edges of the heliostat despite the noisy background, while the DoLP image produced similar results. In the more difficult case for conventional methods where two heliostats overlap, the DoLP image shows a significant advantage over the visible image, producing a clear edge as shown in Fig. 6e compared to the weaker result in Fig. 6d, as the arrows indicate.

In the ground-vs-ground case (Fig. 6f–6g), the AoP image performed better at filtering out background noise such as cracks in the ground and heliostat shadows (arrows in Fig. 6f). However, the second row of heliostat facets failed to appear in the AoP edge detection results. For the other rows, reflections from nearby heliostats caused problems in the visible image (Fig. 6f), while the AoP image showed cleaner facet edges within the heliostat region of interest.

Since corresponding visible images were not available for every polarimetric image collected during the field tests, the accuracy of the method will need to be evaluated in future work. Using the available data and the known shape of the heliostat, we summarize the ability of polarization images to detect edges in each case, as shown in Table 2. Here, successful detection is defined as cases where all facet edges can be identified in the processed polarization images based on the known heliostat facet shape and location.

| Scenario | Successful (No. of sets) | Unsuccessful (No. of sets) | Total (No. of sets) | Success Rate |
|---|---|---|---|---|
| Sky-vs-sky | 108 | 10 | 118 | 91.53% |
| Ground-vs-ground | 29 | 1 | 30 | 96.67% |

**Table 2 Edge detection results summary**

These results suggest that PIHIM improves edge detection in scenarios where conventional imaging struggles, including both sky-vs-sky and ground-vs-ground. This makes it a useful complementary tool for methods like UFACET that require reliable boundary identification. However, the current polarization images have lower resolution, which limits the accuracy of edge-based calculations for optical error evaluation. Combining visible, DoLP, and AoP images may be a promising approach for future work.

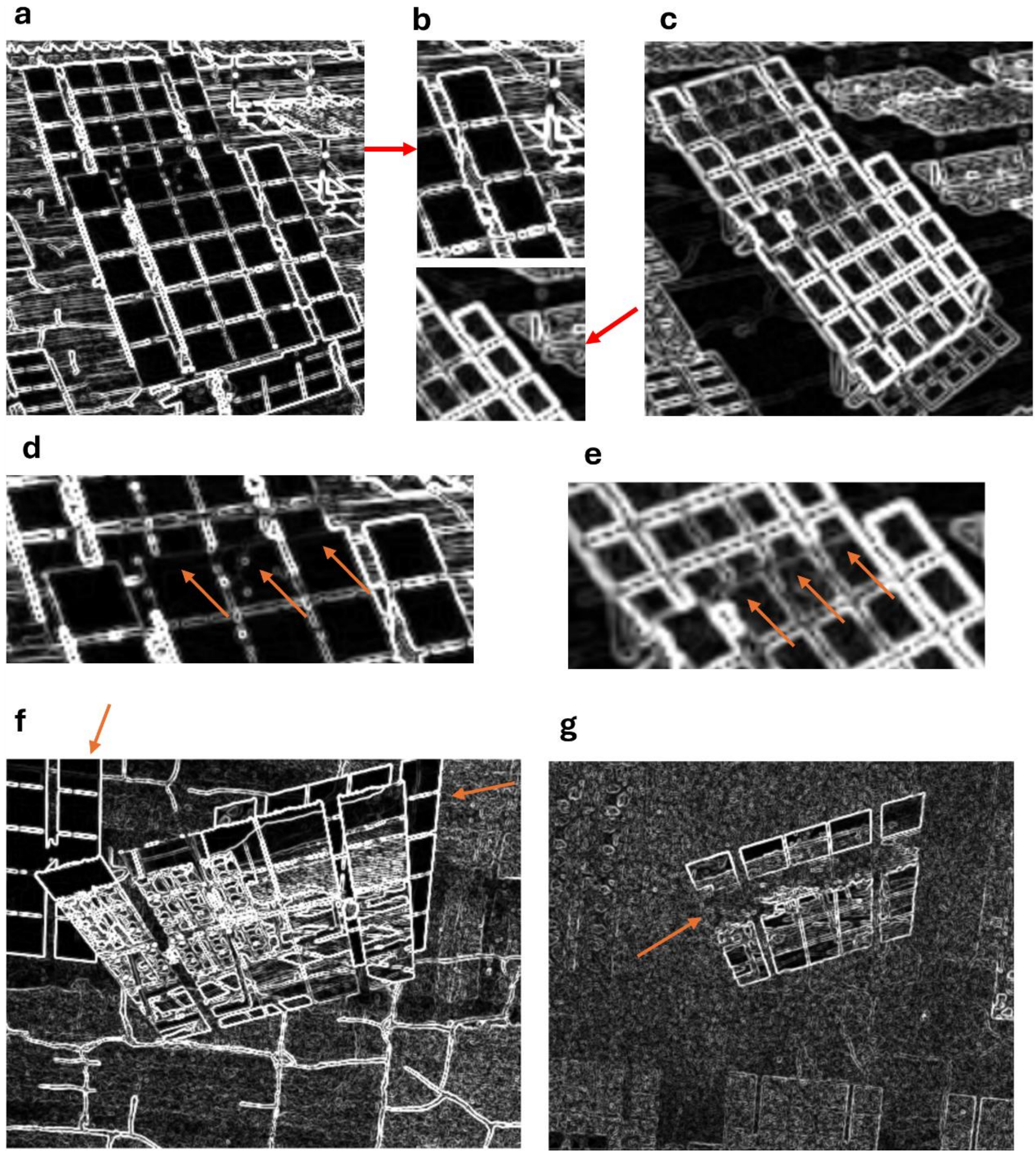

**Figure 6.** Sobel edge detection results applied to visible and polarization images. **a, c** sky-vs-sky case, visible vs. DoLP. **b**, ROIs of both visible and DoLP images showing detectable heliostat edges despite background noise. **d, e,** overlapping heliostat case, visible vs. DoLP, showing the clear advantage of DoLP for edge detection. **f, g**: ground-vs-ground case, visible vs. AoP.

### *4.4. Optical Error Inspection*

Here we present a field test aimed at demonstrating the feasibility of applying polarimetric imaging to heliostat inspection. Because no ground-truth reference data were available, this test should be considered a proof-of-concept study rather than full validation. The primary objective is to illustrate the enhanced contrast achievable with the polarimetric camera under realistic CSP field conditions. Optical error of a heliostat includes the canting error of each facet. To check the effectiveness of our method, we developed a method using similar approaches of the UFACET method to calculate the canting error for each mirror. In PIHIM method, a pair of heliostats were used in this calculation. The images have target heliostat 7E3's reflection in Heliostat Under Assessment (HUA) 8E3. In this study, only polarimetric images were used for quantitative analysis of heliostat edges and crossbar features. The visible images included in figures are provided only for illustration, to demonstrate situations where conventional imaging struggles and where polarimetric imaging offers improved contrast. The edges and the crossbar features are acquired using the polarization image shown in Fig. 7a. The ideal reference canting data for 8E3 shown in Fig. 7b is generated by UFACET software by assuming plano facets and a symmetric paraboloid. In the following sections, the "optical error" calculated are the deviation from this ideal reference data, as the ground truth data of these heliostat facet canting is unknown. Fig. 7c-e show the calculation process. The following procedures are considered in this process:

1. Calibrating lens and camera;
2. Correcting camera position and line-of-sight;
3. Correcting the relative orientation angle between the two heliostats;
4. Calculating canting error for each facet.

Lens distortion and polarization camera transmission efficiency were calibrated in the laboratory before conducting the field test. For lens distortion, a checkerboard calibration target was imaged from multiple viewpoints, and the intrinsic parameters of the camera (focal length, principal point, skew coefficient) were extracted using the OpenCV *calibrateCamera* routine. Both radial (k1, k2, k3) and tangential (p1, p2) distortion terms were included in the pinhole camera model. The calibration error was below 0.4 pixels, which is adequate for UAV-based imaging at the tested distances. Polarization transmission efficiency was also measured for each wire-grid channel (0°, 45°, 90°, 135°) to account for unequal throughput. A correction matrix was applied in preprocessing to normalize the response of all polarization channels, ensuring that AoP and DoLP calculations were not biased by pixel-to-pixel transmission variations.

For the field test, UAV GPS data was used to log the camera's nominal position and orientation. However, the GPS accuracy (~1–2 m) and IMU drift were insufficient for the precision required in canting analysis. Therefore, a photogrammetry-based back-calculation was performed first. We built an optical forward model of the heliostat based on known dimensions of the facets, the focal length and sensor size of the polarization camera, and the approximate UAV position from the flight log. Using this model, synthetic projections of the heliostat corners were generated. The initial UAV pose from GPS was used as a starting guess and then refined by a nonlinear least-squares optimization (Levenberg–Marquardt algorithm). The objective function minimized the root mean square error (RMSE) between the simulated and captured corner locations. This procedure allowed us to refine the UAV pose and <0.2° in orientation relative to the heliostat.

Once the UAV pose was corrected, we accounted for the relative orientation of the target heliostat (7E3) and the heliostat under assessment (HUA, 8E3). The crossbar centers of each facet were chosen as fiducial points, as these features are geometrically well defined and robust in polarization images. The elevation and azimuth angles of both heliostats were extracted from the heliostat operation log provided by the CSP field. These logged values were used as initial conditions, and then adjusted by a rigid-body alignment algorithm until the projected crossbars of the simulated model aligned with those in the captured image. This step was necessary to isolate facet-level canting errors from bulk misalignment between heliostats. The optimization process requires the elevation and azimuth angles of both the HUA and the target heliostat as initial inputs. These values were obtained from the heliostat operation log provided by the CSP field. In total, there are 25 facets on the heliostat, as shown in Fig. 8a. Here, only 16 of the facets shown in Fig. 8a were covered by the reflection of 7E3 and could be used for optical error evaluation, with the data shown in Fig. 8b serving as the reference design data. The mirrors on the top row (facet 1-5) only reflect the sky and the mirrors on column on the right (facet 5, 10, 15, 20, 25) do not contain enough reflection images of 7E3. As we simulate the orientation of the target heliostat changes, we calculate the error of the 16 centers over a range of ±5 degrees, using a step size of 0.1 degrees (~1.75 mrad), until the root mean square error reaches its minimum for both azimuth and zenith angle rotations.

After global orientation correction, facet-level canting errors were computed. For each facet, the simulated reflection of heliostat 7E3 was rotated iteratively in azimuth and zenith over a ±5° range, with a step size of 0.1° (~1.75 mrad). At each step, the simulated facet reflection center was compared with the measured location in the polarization image. The rotation producing the minimum RMSE between simulated and observed facet centers was taken as the best estimate of the canting orientation. This yielded facet-wise canting deviations relative to the UFACET-generated design reference for heliostat 8E3. This way, we can calculate the canting error for each facet with results shown in Fig. 8c and Fig. 8d. The canting deviations from the ideal field on-axis canting angles of heliostat 8E3 are within 10 mrad with a standard deviation of 3.00 mrad for azimuthal angle and 2.11 mrad for zenith angle.

It should be noted that the canting deviations reported here are calculated relative to the ideal design specification of heliostat 8E3 (Fig. 8b), rather than real-world ground-truth canting data. Because the actual canting reference of the heliostat facets was not available during the field tests, we used the design specification as the reference. This provides a useful comparison but does not guarantee alignment with the heliostat's true in-field condition, and the reported deviations should therefore be interpreted as differences from the ideal model rather than absolute canting errors. In practice, measurement accuracy is influenced by factors such as imaging resolution, GPS and camera position accuracy, and heliostat tracking precision. Future work will address these limitations by incorporating ground-truth measurements and improving system accuracy to fully validate the PIHIM method.

PIHIM has to rely on the polarization camera to achieve the level of accuracy for edge position. A low-resolution polarization camera can help a high-resolution camera to determine the edge but with location accuracy limited by the polarization camera resolution. In scenarios where conventional visible imaging cannot resolve heliostat edges (e.g., sky-vs-sky), PIHIM can supply the edge information needed for UFACET or other error detection pipelines. Although full access to UFACET data was not available for direct comparison, the results presented here indicate that polarization contrast would extend the applicability of such methods.

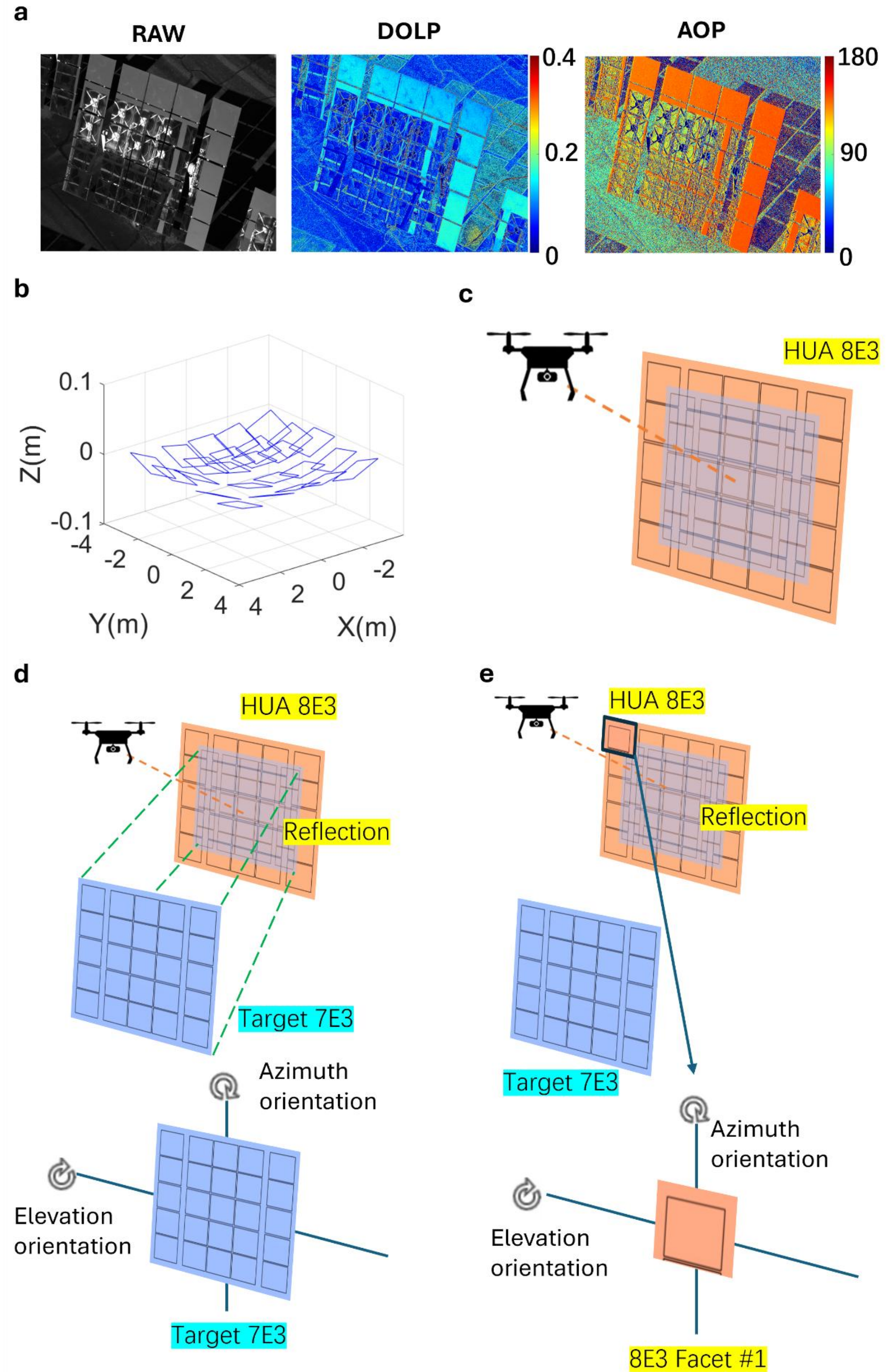

**Figure 7 UFACET optical error calculations with the polarization images on heliostat 8E3 at Sandia NSTTF. a.** The intensity (RAW), DoLP and AoP images taken

with the polarimetric imaging drone on heliostat 8E3 at Sandia NSTTF. **b.** Modeling of the ideal facet canting of 8E3. **c.** Using the corners of the heliostat to perform back-calculations and acquire a more accurate relative camera position. **d.** PIHIM method to find the relative orientation angle between HUA and target heliostat. **e.** PIHIM method to find the canting angle of each facet.

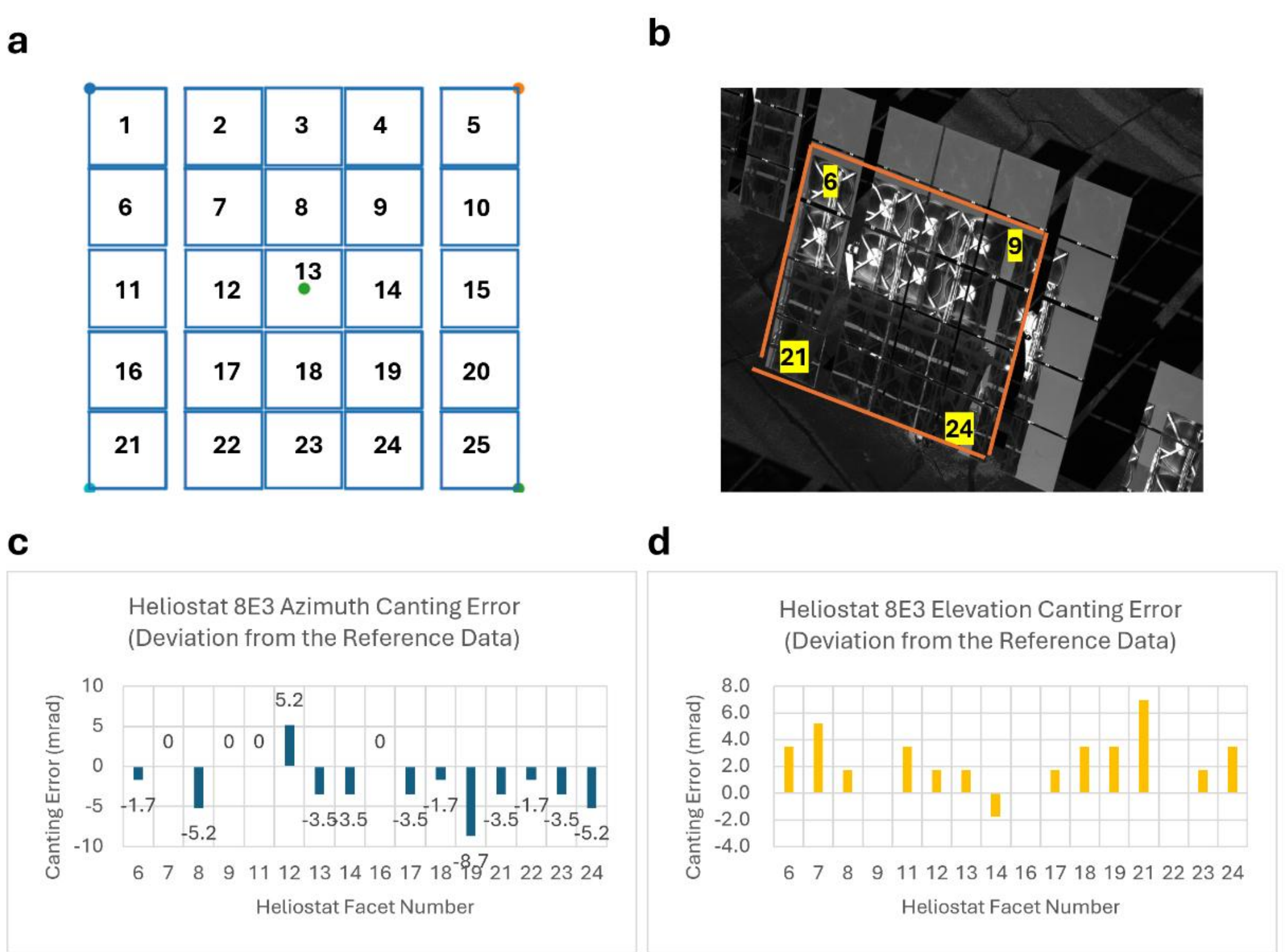

**Figure 8 PIHIM canting deviation calculation results for the bottom 4 rows of facets on heliostat 8E3. a.** Facet number of heliostat at Sandia NSTTF. **b.** The captured images were used to calculate canting deviations for the facets in the orange rectangle. **c.** Azimuth canting error (deviation from the ideal reference data) calculated for facet 6-9, 11-14, 16-19 and 21-24. The facet number with no data points indicates the calculated canting error was zero. **d.** Elevation canting error (deviation from the ideal reference data) calculated for facet 6-9, 11-14, 16-19 and 21-24. The facet number with no data points indicates the calculated canting error was zero.

## 5. Conclusions

The polarimetric imaging system we developed demonstrates strong potential for heliostat inspection in CSP fields. Using a UAV-based scanning approach, inspections can be performed quickly over large areas without interrupting ongoing field operations. Polarimetric imaging can be integrated with conventional visible imaging methods to enhance edge detection and improve accuracy. Compared to traditional UAV-based techniques, the polarimetric approach provides complementary polarization information that facilitates more robust and automated CSP field inspection.

At the same time, several limitations remain. The first one is weather conditions. It is crucial that we use the high DoLP region of the sky for edge detection. During cloudy days or days with high air pollution, more scattering events happen before the incident skylight reaches the heliostat mirror facet, which lowers the overall DoLP. These conditions are not ideal for polarimetric imaging as the contrast will be low. However, considering that solar fields tend to be in areas with significant sunlight exposure, the field operator has the option to wait for a sunny day with clear sky to apply this method. For each field, a techno-economic analysis should be completed beforehand to determine the frequency of field inspection and maintenance to maximize the field efficiency, and the advantage of using UAV-based approach is that it can potentially not be intrusive to the normal tracking of the heliostats. The second constraint is the angle requirement. Depending on the sun's position on the sky dome and the orientation angle of the heliostat, there will be occasions with no good angles for the polarimetric imaging drone to face that simultaneously satisfy the field safety protocols, reflection constraints and the desired high DoLP region requirements. Third, the lower resolution (5M pixels) of currently available polarimetric sensors can limit edge localization accuracy, particularly when only a few pixels define facet boundaries, which propagate into canting error calculations. Other limitations of the current hardware also limit the speed, performance and operation efficiency of optical error detection in the field. Thus, while polarimetric imaging provides clear advantages in terms of contrast enhancement, further improvements in sensor performance are necessary to increase quantitative accuracy.

A further limitation of the present study is the relatively small dataset, which was restricted to a limited number of heliostat pairs and camera positions. In addition, ground-truth canting data were not available during the field tests, which prevented direct quantitative validation against established reference systems such as UFACET. While the current results demonstrate the feasibility of PIHIM, future work should include larger datasets across multiple heliostat configurations, environmental conditions, and reference comparisons to establish full statistical reliability.

Environmental factors such as weather and surface condition also affect PIHIM performance. Cloud coverage and air pollution can cause more scattering of the skylight and thus lower the DoLP of skylight. Mirror soiling and different ground types (e.g., sand, gravel, concrete) add complexity to the reflection simulation model and cause potential inaccuracy in DoLP and AoP simulation. While PIHIM uses relative contrast rather than absolute values, these factors can still reduce edge clarity and should be seen as limits. The impacts of soiling and cloud coverage and methods to use polarimetric imaging system under these conditions are discussed in another work [30]. In PIHIM, we use DoLP as a guideline to judge the feasibility of applying polarimetric imaging for edge detection. When factors such as soiling and cloud coverage are significantly influential, DoLP is largely reduced due to the increased scattering. For this reason, we only process images with DoLP values greater than 0.2 on the heliostat reflection to ensure sufficient DoLP or AoP difference. Future work should test PIHIM under different weather and soiling conditions and apply mitigation strategies such as planning inspections during clear-sky hours, keeping mirrors clean, and adjusting model parameters for local ground reflectance. These steps will improve the reliability of PIHIM for real CSP fields.

In addition, the primary sources of error in PIHIM stem from the quality of the captured data, specifically whether the UAV is positioned at angles that maximize DoLP or AoP contrast. When optimal viewing geometry is not achieved, edge detection quality and subsequent canting error calculation are degraded. At the time of the field tests, we did not have access to the same data as used in UFACET or other optical error evaluation methods. For this reason, we developed our own calculation procedures, which are not directly or quantitatively comparable with existing references. Future developments should therefore include benchmarking PIHIM with established methods such as UFACET or other ground-truth validated systems. We emphasize that PIHIM is not intended to replace existing heliostat inspection methods, but rather to complement them. Any approach that currently relies on conventional imaging can be adapted to a combined conventional and polarimetric imaging system, thereby enhancing performance in scenarios where conventional imaging alone faces difficulty (e.g., sky-vs-sky and ground-vs-ground cases).

In our current field tests, a typical 20-minute UAV flight allowed imaging of six heliostat pairs (~12 heliostats), corresponding to ~36 heliostats per hour per drone. With anticipated upgrades, including faster image acquisition, improved flight planning, and longer endurance, each drone is expected to scan 50–100 heliostats per hour. For a 10,000-heliostat field, this would reduce inspection time to 100–200 drone-hours, meaning that with a fleet of ~10 drones, full inspection could be achieved in less than a single day. Scaling efficiency is therefore determined primarily by UAV endurance, coordinated fleet operation, and optimized path design. Importantly, the method does not interrupt field operation, though optimal inspection times are influenced by solar position.

As our work focused on heliostat fields, the same method and system can also be applied to other inspection tasks with minimal modifications on the modeling, such as parabolic trough plants. With the current limitations mainly falling on the specification of the polarization camera models, we expect that more advanced polarization cameras with better resolution and acquisition speed can become more commercially attractive in the clean energy field in the near future. Besides optical error inspection, polarimetric imaging also provides other advantages such as crack detection [1] and mirror soiling detection [30, 31] for heliostats.

## 6. CRediT authorship contribution statement

**Mo Tian**: Conceptualization, Data Curation, Formal analysis, Investigation, Methodology, Software, Validation, Visualization, Writing – original draft, Writing – review & editing, System development, Data collection. **Kolappan**

**Chidambaranathan**: Software, System development, Data collection. **Md Zubair Ebne Rafique**: Formal analysis, Visualization. **Neel Desai**: Software, System development, Data collection. **Jing Bai**: Conceptualization, Methodology, Experiment design, Data Collection. **Randy Brost**: Investigation, Resources, Supervision, Project administration, Data collection, Writing – review & editing. **Daniel Small**: Software, System development, Data collection. **David Novick**: Software, System development, Data collection. **Julius Yellowhair**: Conceptualization, Methodology, Experiment design. **Yu Yao**: Conceptualization, Formal analysis, Funding Acquisition, Investigation, Methodology, Project administration, Resources, Supervision, Writing – review & editing.

## 7. Funding

The research conducted at Arizona State University is supported by the U.S. Department of Energy Solar Energy Technology Office under the contract no. EE0008999. Sandia National Laboratories is a multi-mission laboratory managed and operated by National Technology & Engineering Solutions of Sandia, LLC, a wholly owned subsidiary of Honeywell International Inc., for the U.S. Department of Energy's National Nuclear Security Administration under contract DE-NA0003525.

## 8. Data Availability

The data supporting the findings of this study are available from the authors upon reasonable request.

## 9. Acknowledgements

The authors gratefully acknowledge Anthony Evans, Kevin Good, Kevin Hoyt, and George Slad for their assistance with data collection during the heliostat field and flight operations at Sandia NSTTF.

## 10. Nomenclature

| **Symbol/Abbreviation** | **Definition** |
|---|---|
| AoP | Angle of Polarization (°) |
| CSP | Concentrating Solar Power |
| DoLP | Degree of Linear Polarization |
| HUA | Heliostat Under Assessment |
| NSTTF | National Solar Thermal Test Facility |
| NIO | Non-Intrusive Optical method |
| PIHIM | Polarimetric Imaging Heliostat Inspection Method |
| RMSE | Root Mean Square Error |
| UFACET | Universal Field Assessment, Correction, and Enhancement Tool |
| UAV | Unmanned Aerial Vehicle |
| I | Intensity of light |
| Q | Stokes parameter (linear polarization, 0°/90°) |
| U | Stokes parameter (linear polarization, 45°/135°) |
| V | Stokes parameter (circular polarization, not used) |
| θ | Zenith angle (°) |
| φ | Azimuth angle (°) |
| ΔAoP | Difference in AoP between reflected and actual ground regions |
| ΔDoLP | Difference in DoLP between adjacent heliostat reflections |
| σ | Standard deviation |

## 11. Declaration of Generative AI and AI-assisted Technologies in the Writing Process

During the preparation of this work, the authors used ChatGPT-5 solely for grammar, spelling, and formatting checks. After using this tool, the authors reviewed and edited the content as needed and take full responsibility for the content of the publication. No original scientific content was generated using AI tools.

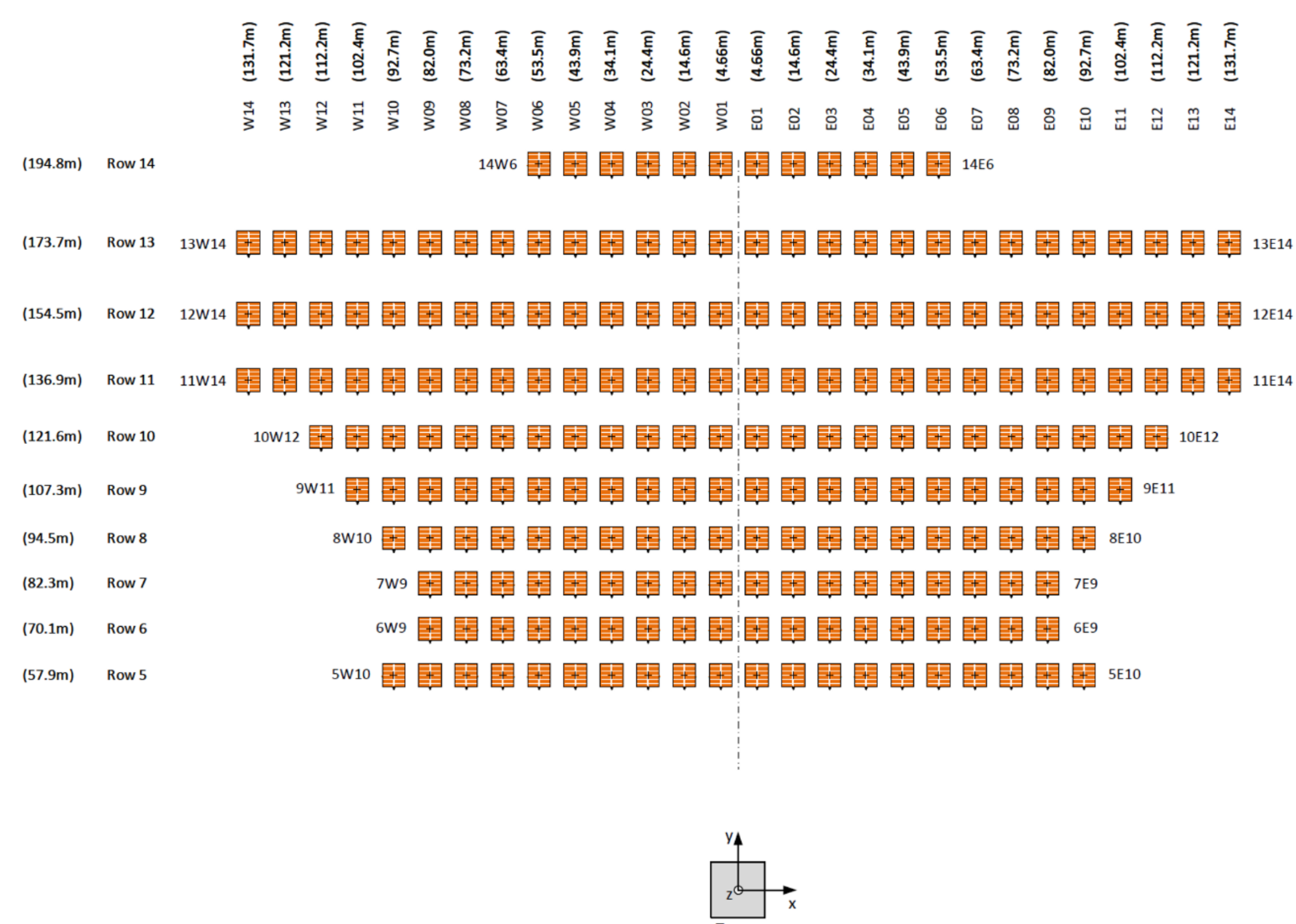

**Supplementary Figure S1. Sandia NSTTF field coordinates**. Each heliostat is named as “row, column”. For example, 14E3 is the heliostat on the 14$^{th}$ Row and 3$^{rd}$ East from the center y axis.

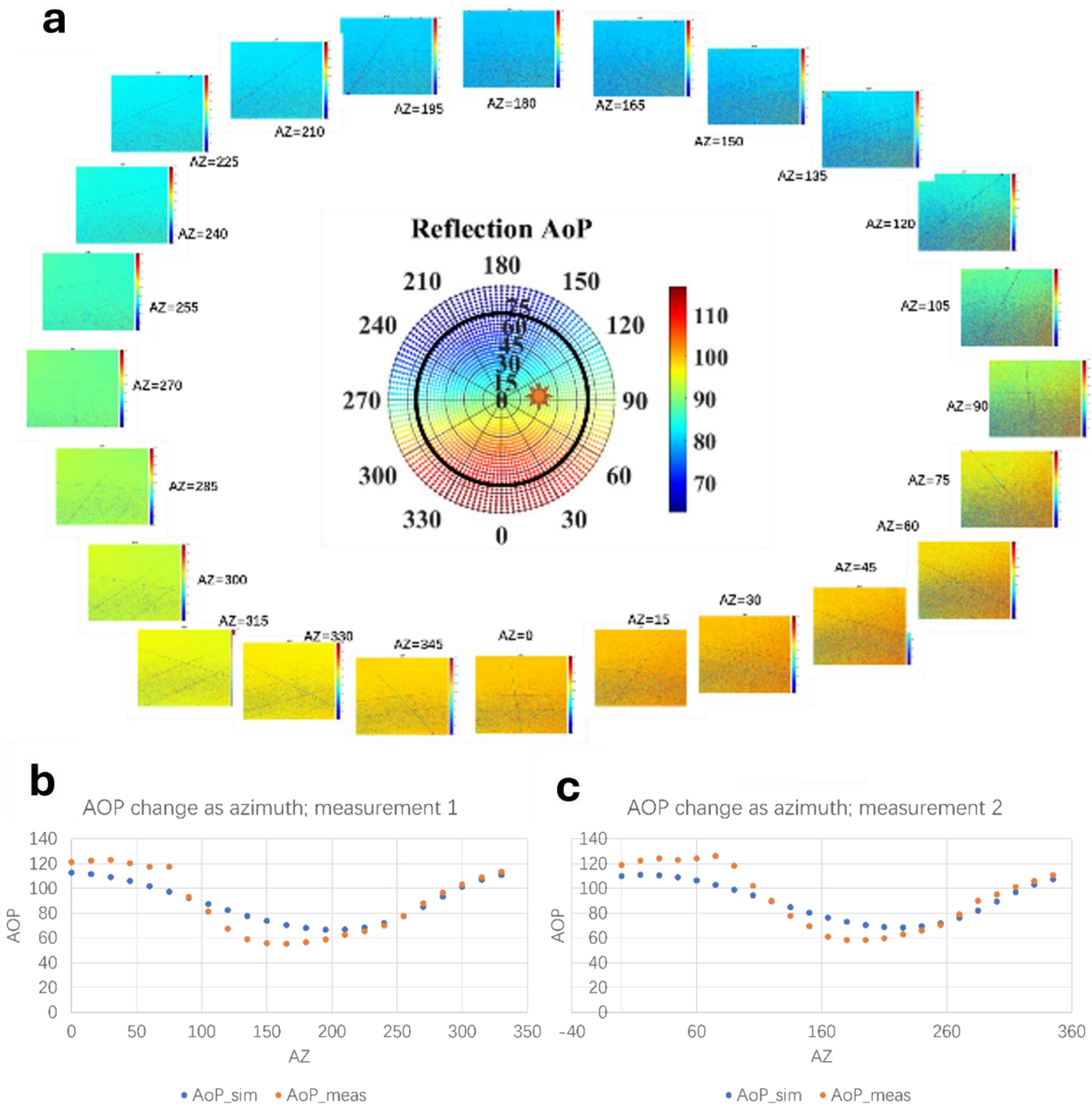

**Supplementary Figure S2. a,** Simulation of the reflected skylight AoP mapping and the corresponding measurement AoP image at each angle. **b-c,** comparison of simulated and measured Angle of Polarization (AoP) across different azimuthal angles at two time points. Orange dots represent the averaged measured AoP values, while blue dots show the corresponding simulation results. The observed alignment between the two sets confirms that the AoP trends consistently follow the azimuthal angle, demonstrating good agreement between measurement and simulation.